\documentstyle[preprint,aps]{revtex}
\def\dsp{\def\baselinestretch{1.3}\large\normalsize}
\begin{document}
\preprint{ANL-1999}
\draft
\def\nuc#1#2{$^{#1}{\rm {#2}}$}
\def\overlay#1#2{\setbox0=\hbox{#1}\setbox1=\hbox to \wd0{\hss #2\hss}#1%
\hskip -2\wd0\copy1}

\dsp{
\title{Fission Hindrance in hot $^{216}$Th: Evaporation Residue Measurements}

\author{    B.\ B. Back, D. J. Blumenthal, C. N. Davids, D. J. Henderson,\\
    R. Hermann, D. J. Hofman, C. L. Jiang, H. T. Penttil\"{a}, and A. H. Wuosmaa}

\address{ Argonne National Laboratory, Argonne, IL 60439, USA}
\date{\today}
\maketitle
\begin{abstract}
The fusion evaporation-residue cross sections for $^{32}$S+$^{184}$W have
been measured at beam energies of E$_{beam}$= 165, 174, 185, 196, 205, 215,
225, 236, 246, and 257 MeV using the ATLAS Fragment Mass Analyzer. The data
are compared with Statistical Model calculations and it is found that a
nuclear dissipation strength, which increases with the excitation energy is 
required to reproduce the excitation function. A comparison with previously
published data shows that the dissipation strength depends strongly on
the shell structure of the nuclear system.
\end{abstract}


\section{Introduction}

Experimental studies of the time-scale of fission of hot nuclei have recently
been carried out using the emission rates of neutrons \cite{Hinde}, 
$\gamma$-rays \cite{Hofman}, and charged particles \cite{Lestone} as "clocks"
for the fission process. These experiments have shown that the fission 
process is strongly hindered relative to expectations
based on the statistical model description of the process. The observed
effects extend well beyond any 
uncertainties in the model parameters. It therefore appears that a dynamical 
description of the fission process at these energies is more appropriate
\cite{Froebrich} and that the experimental data are able to shed light 
on dissipation effects in the shape degree of freedom. However, these experiments
are not very sensitive to whether the emission occurs mainly before or 
after the traversal of the saddle point as the system proceeds toward scission.
Various dissipation models are, however, strongly dependent on the deformation
and shape symmetry of the system. As an alternative to these methods we
therefore measure the evaporation probability for hot nuclei formed in 
heavy-ion fusion reactions, which is sensitive only to the dissipation strength
inside the fission barrier. 
As the hot system cools down by the emission of neutrons
and charged particles there is a finite chance to undergo fission after each
evaporation step. If the fission branch is suppressed due to dissipation
there is therefore a strongly enhanced probability for survival which 
manifests itself as an evaporation residue cross section which is larger than
expected from statistical model predictions. This effect depends, however, only 
on the dissipation strength inside the saddle point and may therefore
provide the desired separation between pre-saddle and post-saddle dissipation.

In this paper, we report on recent measurements of evaporation residue cross 
sections for the $^{32}$S+$^{184}$W system over a wide range of beam 
energies using
the Argonne Fragment Mass Analyzer (FMA). In sect. II we describe the
experimental procedure followed by a discussion of the measurements of
absolute evaporation residue cross section in sect. III. The results are
compared to statistical model calculations and other relevant data in sect. IV
followed by the conclusion, sect. V.

\section{Experimental arrangement}

The measurements were carried out using $^{32}$S-beams from the ATLAS
superconducting linac at Argonne National Laboratory.
The cross sections for evaporation residues produced in the
$^{32}$S+$^{184}$W reaction were measured at beam energies of 165, 174, 185,
195, 205, 215, 225, 236, 246, and 257 MeV. Targets of isopically separated
$^{184}$W with thickness
200~$\mu$g/cm$^2$ on a 100~$\mu$g/cm$^2$ carbon backing were used. 
The Argonne Fragment Mass Analyzer \cite{Davids} was used for identification
of evaporation residues. A schematic illustration of the setup
is shown in Fig. 1.

In these experiments a sliding seal target chamber
was used, which  allows for measurements at angles away from 0$^\circ$.
This is required in order to obtain the angular
distributions for integration of the total evaporation residue cross section.
Elastically scattered S-ions were registered in a Si detector placed at
30$^\circ$
relative to the beam axis with a solid angle of $\Omega_{mon}$ = 0.249 msr.
These data were used for normalization purposes.
A~ 40 $\mu$g/cm$^2$ carbon foil was placed 10 cm downstream from the target
to reset the charge state of 
reaction products, which may be abnormally highly charged as a result of
Auger electron emission following the $\gamma$ decay of short-lived isomers.

A square entrance aperture for the FMA covering $\theta,\phi
=4.5^\circ\times4.5^\circ$
($\Omega_{FMA} = 6.24$ msr) was used.
Reaction products transmitted through the FMA were dispersed in M/q
(mass/charge) at the focal plane, where the spatial distribution was measured
by a thin x-y position sensitive avalanche detector. When the FMA was placed
at 0$^\circ$ some settings of the electrostatic and magnetic fields of the
instrument allows beam particles scattered off the anode of the first
electrostatic dipole, ED1,
to be transported to the focal plane (presumably after
a subsequent forward scattering in the vacuum chamber of the magnetic dipole
MD1). When measuring small cross sections, as in the present study, it is
therefore mandatory to achieve a clean separation between evaporation
residues and beam particles. This was achieved by measuring their flight-time
over the  40~cm distance to a double-sided Si strip detector (DSSD)
placed behind the focal plane. This detector has a total active area of
5$\times$5 cm$^2$ and is divided into 16 strips on both the front and rear
surface arranged orthogonally to each other.
The information on the particle mass obtained from the
time-of-flight and energy measurement provided by the Si-detector gave a
clean discrimination against the scattered beam as illustrated in Fig. 2.
The efficiency for transporting evaporation residues from the focal plane
to the Si-detector was determined from the spatial distribution over the face
of the DSSD detector as shown in Fig. 3 for these beam energies. By Gaussian
extrapolation of the distribution beyond the edge of the detector it is
estimated that this efficiency is around $\epsilon_{PPAC-Si}$ = 87\%. 

The transport efficiency of the FMA as a function of the mass, energy amd
charge-state of the ion has been determined in a separate experiment
\cite{Back}.

\section{Cross sections}

The evaporation residue cross section for the $^{32}$S+$^{184}$W reaction 
was measured for beam energies in the range $E_{beam}$=165-257 MeV. 
Evaporation residues were identified by time-of-flight and energy measurement
using the focal plane PPAC detector and the Si-strip detector placed 
ca.~40~cm behind the focal plane. 
The charge 
state distributions, which were measured at three beam energies, are shown
in Fig. 4. The dashed curves represent the 
formula of Shima {\it et al.} \cite {Shima}, whereas a somewhat better fit
to the data is given by the Gaussian fit (solid curves) with a fixed standard
deviation of $\sigma$=3. The arrows indicate the charge state setting of the 
FMA used for the cross section measurement. The derivation of the evaporation
residue cross section at intermediate beam energies is based on an interpolation
between these measured charge state distributions.

Since the FMA disperses in $M/q$ at the focal plane there will be cases of 
ambiguities in the mass identification, since overlaps between lighter mass 
products from one charge state, $q$, will invariably overlap with heavier
products from the neighboring charge state, $q+1$, when compound nuclei with
high excitation energy are studied, see Fig. 5. We are not able to  
resolve this ambiguity with the present setup, and have therefore obtained the
cross sections by integrating all counts that fall between the positions for
$M/(q-\frac{1}{2})$ and $M/(q+\frac{1}{2})$ along the focal plane. Since the 
FMA is set up for the most abundant charge state, $q$ and mass, $M$, we 
expect that the loss of residues with charge state, $q$, and masses that fall
outside this window is compensated by the acceptance of  residues with 
charge states, $q+1$ and $q-1$ that fall inside this window.

\subsection{Detection efficiency}

The transport efficiency as a function of recoil
energy and mass relative to the setting for the FMA
has been measured for
monoenergetic particles by observing the recoils from
elastic scattering of $^{32}$S + $^{197}$Au, $^{208}$Pb, $^{232}$Th \cite{Back}.
To correctly estimate the transport efficiency for evaporation
residues, which have an extended energy distribution, it is necessary to
fold the energy distribution with the measured acceptance curve. The energy
distribution was not measured directly in the present experiment, but the yield
of residues as a function of the energy setting of the FMA was measured as
shown in Fig. 6 (top panel). In principle, since the energy acceptance of
the FMA is known, it should be possible to convert this measurement into an
energy distribution with some accuracy.

We have, however, used a slightly different method which incorporated both
this measurement and the measurement of the angular distributions.
Assuming that both the angular distribution of evaporation residues and their
energy distribution at 0$^\circ$ arise from isotropic multiparticle emission
from the hot compound nucleus, these two entities are related by the
kinematics of the particle decay cascade. We assume that the 
recoil energy distribution is isotropic in the center-of-mass system and
that it has a Maxwellian form, namely

\begin{equation}
\frac{dP}{dE_{cm}} =
\frac{2}{\sqrt{\pi}}\frac{\sqrt{E_{cm}}}{a^{3/2}}\;
\exp\left(-E_{cm}/a\right),
\end{equation}
where $E_{cm}$ is the recoil energy in the center-of-mass system and
$a = \frac{2}{3}\langle E_{cm}\rangle$ is two
thirds of its average value. The energy distribution in the laboratory
system at $\theta = 0^\circ$ is then 

\begin{equation}
\frac{dP}{dE_{lab}}|_{_{0^\circ}} =
\frac{1}{2} \left( \frac{1}{\pi a} \right)^{3/2} 
\sqrt{E_{lab}}\; \exp\left[
-\left(\sqrt{E_{lab}}-\sqrt{E_{CN}}\right)^2/a\right].
\end{equation}
Here, $E_{CN}$ is the laboratory energy of the compound nucleus prior to the
particle evaporation cascade. A small correction to $E_{CN}$ arising from the 
mass loss due to particle evaporation has been ignored in eqs. 1-3.
Similarly we find the angular distribution

\begin{equation}
\frac{dP}{d\Omega_{lab}} =
\frac{1}{2} \left( \frac{1}{\pi a}\right)^{3/2} \int_0^{\infty}
\sqrt{E_{lab}}\;
\exp\left[-\left(E_{lab}+E_{CN}-2\sqrt{E_{lab}E_{CN}}\cos\theta\right)/a\right]\;dE_{lab}.
\end{equation}

We find that a value of $a$ = 0.5 MeV gives a good representation of both
the transmission as a function of the energy setting of the FMA, $E_{FMA}$
and the measured angular distribution, see Fig. 6. For the angular
distribution we have also taken the effects of multiple scattering in the
target and backing material as well as the charge state reset foil into
account. This increases the width of the angular distribution somewhat  and
results in good agreement with the data as
shown by the solid curve in Fig. 6b. This value of $a$ = 0.5 MeV
corresponds to a transport efficiency of the FMA for evaporation residues
of $\overline{\epsilon_{FMA}} \approx 0.60$, see Table I.

\subsection{Angular distributions}

The angular distributions of evaporation residues were measured at three
beam energies utilizing
the sliding-seal target chamber for the FMA. Differential cross sections,
$d\sigma/d\Omega$, as 
a function of the mean angle, $<\theta_{lab}>$, relative to the beam axis are
shown  in the left side panel of Fig. 7. The right side panel shows the
cross sections converted to $d\sigma/d\theta$, which is relevant for the angular
integration of the total evaporation residue cross section. 
The angle 
integrated cross sections are thus derived from a fit to the data
expressed in terms of $d\sigma/d\theta$ using the function 
$2\pi\sin\theta dP/d\Omega_{lab}$. The curves shown in the left 
side panel of Fig. 7 are computed by removing the $2\pi\sin\theta$ term. We
observe that these latter curves underrepresent the differential cross section 
at small angles indicating that the angular distribution really has two
components. However, we do not feel that the data are of sufficient quality
to allow for a reliable separation of two components and by observing the fits
to the $d\sigma/d\theta$ data it is clear that only a very small error could
arise from this simplification.

The data shown in Fig. 7 are corrected for the efficiency for
transporting evaporation residues through the FMA. We estimate this transport
efficiency, $\epsilon_{FMA}$,  by folding the
energy distribution of the evaporation residues with the energy acceptance
of the FMA, which was measured by Back {\it et al} \cite{Back} for the
entrance aperture used in this experiment. The mean energy of the
compound system, $E'_{CN}$ ( corrected for the energy losses in the target material, backing and the reset
foil ), is determined from the reaction kinematics and
listed in Table I. The parameter, $a =
\frac{2}{3}\langle E_{cm}\rangle$ was determined to have a value of about
$a \approx 0.5$  for the $E_{beam} = $ 246 MeV point by simultaneously
fitting the angular distribution of evaporation residues and a scan of the
energy setting of the FMA, see Fig. 6. The value of $a$  was for the other
beam energies 
scaled according to $a \propto \sqrt{E^*}/22.4$, which was found to reproduce
also the angular distributions measured at $E_{beam}$ = 174 and 205 MeV, see
Fig. 7.

\subsection{Total evaporation residue cross sections}

The total evaporation residue cross section, $\sigma_{ER}$, is obtained from
the measurement of the differential cross section at $\theta = 5^\circ$,
which was performed at all beam energies. The ratio, $f(E_{beam})  =
\sigma_{ER}/\frac{d\sigma_{ER}}{d\theta}(5^\circ)$, of the angle integrated
cross section, $\sigma_{ER}$, to the measured differential cross section,
$\frac{d\sigma_{ER}}{d\theta}(5^\circ)$, is obtained by smooth interpolation
between the values of $f(E_{beam})$ = 0.089, 0.088, and 0.086 rad obtained
from the angular distribution measurements at $E_{beam}$ = 174, 205, and 246
MeV, respectively.  The total evaporation residue cross sections are then
given by
\begin{eqnarray}
\sigma_{ER}& =& f(E_{beam}) \frac{d\sigma(5^\circ)}{d\theta} \\
            &=& f(E_{beam})\frac{N_{ER}(5^\circ)}{N_{mon}}\;
	      \frac{\Omega_{mon}}{\Omega_{FMA}}\;
2\pi \sin(5^\circ)\frac{d\sigma_{Ruth}}{d\Omega}(30^\circ)\;
\frac{1}{\epsilon_{FMA} \epsilon_{PPAC-Si} P(q)}\nonumber
\end{eqnarray}
where, $N_{ER}(5^\circ), N_{mon}$ are the number of evaporation residue
counts observed at the FMA focal plane Si-detector, and the number of
elastically scattered $^{32}$S ions registered in the monitor detector,
respectively. The differential Rutherford cross section in the laboratory
system is denoted $d\sigma_{Ruth}/d\Omega$, and
$P(q)$ is the fraction of evaporation residues in the
charge state, $q$, for which the FMA was tuned. The charge state fraction,
$P(q)$, was obtained by interpolation of the central charge state, $q_0$
resulting from the fits to the measured distributions with a Gaussian with a
standard deviation of $\sigma$ = 3 charge state units. 

The resulting evaporation residue cross sections for the $^{32}$S+$^{184}$W 
reaction are shown as filled circles in Fig. 8 and listed in Table I. The
measurements are assigned a systematic error of 20\%, mainly due to the
procedure for estimating the transport efficiency through the FMA. 

Fission-like cross sections
and a derived estimate of the complete fusion cross sections for 
the $^{32}$S+$^{182}$W reaction are shown as open circles \cite{Glagola,Keller}
and open squares \cite{Keller}, respectively, along with theoretical 
calculations using a modified Extra Push model \cite{Toke}. 

\section{Comparison with statistical model calculations}

In Fig. 8, the evaporation residue data are
compared with a statistical model calculation obtained with the code CASCADE
(long dashed curve labeled $\gamma$=0) 
using Sierk fission barriers \cite{Sierk} scaled by a factor of
0.9 to approximately account for the cross section at low beam energies, and
using level density parameters of $a_n = a_f = A/8.8$ MeV$^{-1}$.
We observe that the measured cross section increases with beam energy, whereas
the statistical model predicts a decreasing cross section because of an increased
probability for fission during the longer evaporation cascades. 
For comparison we have also performed CASCADE calculations 
using level density parameters of $a_n = A/8.68$ MeV$^{-1}$ and
$a_f = A/8.49$ MeV$^{-1}$ as suggested by T\={o}ke and Swiatecki
\cite{Toke81} (dotted curve),  and $a_n = A/11.26$ MeV$^{-1}$ and 
$a_f = A/11.15$ MeV$^{-1}$ by Ignatyuk {\it et al} \cite{Ignatyuk}
(dotted-dashed curve). Using these values results in an
even sharper decrease of the predicted evaporation residue cross section
with beam energy as shown in Fig. 8. This is a consequence of the fact that
the fission decay rate increases more rapidly with excitation energy when
values of $a_f > a_n$ are used. Although it is expected that $a_f > a_n$ on 
rather firm theoretical grounds we have, however, used the standard values
of $a_f =  a_n = a/8.8$ in order to be able to compare to other works, 
where this value was used in the analysis.  

We hypothesize that the observed increase  of the measured evaporation residue
cross section with excitation energy, which is at variance with 
the statistical model calculations, can be attributed to
an increased hindrance of the fission motion with excitation energy. Fission
hindrance at high excitation has previously been shown to explain
observations of enhanced emission of pre-scission neutrons \cite{Hinde},
charged particles \cite{Lestone}, and $\gamma$-rays \cite{Hofman}, as well
as recent observation of an enhanced survival probability of excited target
recoils from deep inelastic scattering reactions \cite{Hofman98}.

The inclusion of friction in the fission motion results in a modification of
the normal Bohr - Wheeler expression \cite{Bohr-Wheeler}
for the fission decay width, $\Gamma_f^{BW}$  as pointed out by 
Kramers \cite{Kramers}, {\it i.e. }
\begin{equation}
\Gamma_f^{Kramers} = \Gamma_f^{BW} (\sqrt{1+\gamma^2} - \gamma)[1-\exp(-t/\tau_f)]
\end{equation}
where $\gamma = \beta/2\omega_0$ is a reduced nuclear friction coefficient, and $\tau_f$ is a
charactistic time for the building of the fission flux over the saddle point.
$\beta$ denotes the reduced dissipation constant and $\omega_0$ describes
the potential curvature at the fission saddle point. The modification to the
Bohr-Wheeler expression for the fission width thus consists of an overall
reduction given by the so-called Kramers factor, $\sqrt{1+\gamma^2} - \gamma$,
as well as a time dependent in-growth of the fission rate given by the
factor $1-\exp(-t/\tau_f)$ \cite{Grange}. These modifications 
to the fission decay width
has been incorporated into the CASCADE statistical model code in an
approximate way \cite{Butsch}, which has, however, been shown \cite{Back93}
to be very accurate over the applied range of parameters.

Because the 
evaporation residue cross section is such a small fraction of the complete
fusion cross section we find that it is very sensitive to the 
nuclear viscosity of the system inside the barrier. The thin solid curve 
in Fig. 8 represent
a statistical model calculation  where the effects of viscosity are included 
using a linear normalized dissipation coefficient of $\gamma$=5, corresponding 
to a strongly overdamped motion in the fission degree of freedom. This is
approximately the dissipation strength expected from the one-body
dissipation mechanism \cite{Blocki}.
We see that 
this leads to an increase of about a factor 10-20 in the evaporation residue 
cross section relative to the pure statistical model estimate (long dashed
curve), but the overall shape of the excitation function is virtually
unchanged. Within this framework it therefore appears that the viscosity (or 
dissipation) increases rather rapidly over this range of beam energies i.e.
from 200 to 260 MeV, which corresponds to an excitation energy range of
$E_{exc}$=85-136 MeV. Similar effects have been observed in studies of
pre-scission $\gamma$-rays \cite{Hofman} albeit in that case it appears to 
take place over an even smaller excitation energy interval.

In order to deduce the temperature dependence of the dissipation strength in
the fission degree of freedom $\gamma(T)$ that reproduces the observed
increase of the evaporation residue cross section, we have performed a
series of calculations at each beam energy, varying the value of $\gamma$ to
reproduce the measured cross section. This procedure leads to the thick solid
cross section curve going through the data points in Fig. 8; the
corresponding values of $\gamma(T)$ are plotted as solid triangles in Fig. 9.
Note that there is some inconsistency in this approach because the value of
the dissipation strength is ${\it not}$ allowed to vary as the system cools
down during the particle evaporation cascade.
Rather, the dissipation strength is kept constant
throughout the cascade with the value needed to fit the measured evaporation
residue cross section for this particular beam energy. Although this has
been recognized as a shortcoming of these calculations, we have employed this
procedure to be able to compare to other published data analyzed in the same
way.

\section{Discussion}

The dissipation strength in the fission process has recently been measured
by several methods, and it is of interest to compare these different results.
In Fig. 9 we show the normalized dissipation strength parameter, $\gamma$,
obtained from the analysis of 1) the survival probability of Th-like nuclei
excited in deep-inelastic scattering reaction of 400 MeV $^{40}$Ar+$^{232}$Th
\cite{Hofman98} (solid squares), 2) the evaporation residue cross section and
pre-scission $\gamma$-ray emission from the $^{16}$O+$^{208}$Pb
\cite{Brinkmann,Hofman96} (solid diamonds), 3) the present data (solid
triangles), and 4) the fission cross section for $^3$He+$^{208}$Pb reaction
\cite{Rubehn,Back98} (open circles). We observe that the dissipation strength
required to reproduce the different data falls into two groups, namely one
which increases rather sharply above an excitation energy of $E_{exc} \sim$
40 MeV, and another group that increases slowly only above
$E_{exc} \geq$ 80 MeV. 

It is interesting to note that this behaviour may be
related to the shell structure of the compound system. The two systems that
have a closed (or nearly closed) neutron shell at N=126 show only moderate
fission dissipation strength up to high excitation energy, whereas the
mid-shell systems with N = 134 , 142 display a strong increase in $\gamma$
above  $E_{exc} \sim$ 40 MeV. 

Recently, there has been much theoretical interest in the study of the
dynamics of the fission process, both in terms of the description of
experimental observables on the basis of phenomelogical assumptions of the
dissipation strength
\cite{Froebrich,Dhara}
as well as more fundamental theories for the dissipation mechanism itself
\cite{Hofmann,Kolomietz,Magierski,Mukhopadhyay}.
Although the overall dissipation strength found to reproduce the present
data is in fair agreement with estimated based on the simple one-body
dissipation model, namely $\gamma \approx 5-6$ the rather striking increase
with excitation energy (or temperature) is unexplained within this
mechanism, which has no temperature dependence. It is interesting to note
that the linear response theory approach \cite{Hofmann} appears to predict
the increase in dissipation strength although the present development level 
of this theory is not directly applicable for comparison with the
experimental data.

\section{Conclusion}

Measurements of evaporation residue cross sections for heavy fissile systems
are shown to provide rather direct evidence for the fission hindrance 
(or retardation) which is caused by strong nuclear dissipation in the fission
degree of freedom for hot nuclei. The data obtained for the $^{32}$S+$^{184}$W
system show an increasing evaporation cross section with excitation energy,
whereas a decrease is expected on the basis of statistical model considerations
and calculations. The data indicate an increase in the linear normalized
dissipation coefficient $\gamma$ from $\gamma$=0 at $E_{exc}$=85 MeV to 
$\gamma$=5 at $E_{exc}$=135 MeV. Although hints of such an increase have been
obtained within the framework of linear response theory, no direct
comparison can be made with the experimental data. Further study,
both experimental and theoretical, of this phenomenon is warrented, 

This work was supported by the U. S. Department of Energy, Nuclear Physics
Division, under contact No. W-31-109-ENG-38.

\newpage
\begin{table}
\caption{Reaction parameters and total evaporation residue cross section.
$E'_{CN}$ is the kinetic energy of the fused $^{216}$Th system corrected for
energy losses in the target material, backing and the charge state
reset foil,
$A_{ER}$ is the average mass of the evaporation residue,
$E_{FMA}$ is the energy setting of the FMA,
$\overline{\epsilon_{FMA}}$ is the resulting average transport efficiency
through the FMA for the assumed distribution of evaporation residues, and
$\sigma_{ER}$ is the total evaporation residue cross section.}
\label{Table1}
\begin{tabular}{ccccccc}
$E_{beam}$   &   $E'_{CN}$ & $A_{ER}$ & $a$ & $E_{FMA}$ &
$\overline{\epsilon_{FMA}}$ & $\sigma_{ER}$\\
(MeV)  &  (MeV)  & (u) & (MeV) & (MeV) & & ($\mu$b) \\
\tableline
165 & 19.3 & 212 & 0.34 & 22 & 0.61 & 72 $\pm$ 14\\
174 & 20.4 & 212 & 0.36 & 23 & 0.60 & 47 $\pm$ 10\\
185 & 21.7 & 211 & 0.38 & 25 & 0.61 & 63 $\pm$ 13\\
196 & 23.0 & 210 & 0.41 & 26 & 0.60 & 68 $\pm$ 14\\
205 & 24.1 & 209 & 0.43 & 27 & 0.60 & 75 $\pm$ 15\\
215 & 25.3 & 208 & 0.44 & 29 & 0.61 & 103 $\pm$ 20\\
225 & 26.5 & 207 & 0.46 & 30 & 0.61 & 115 $\pm$ 23\\
236 & 27.9 & 206 & 0.48 & 31 & 0.60 & 125 $\pm$ 25\\
246 & 29.1 & 205 & 0.50 & 33 & 0.61 & 190 $\pm$ 38\\
257 & 30.4 & 204 & 0.52 & 35 & 0.60 & 175 $\pm$ 35\\
\end{tabular}
\end{table}

\newpage

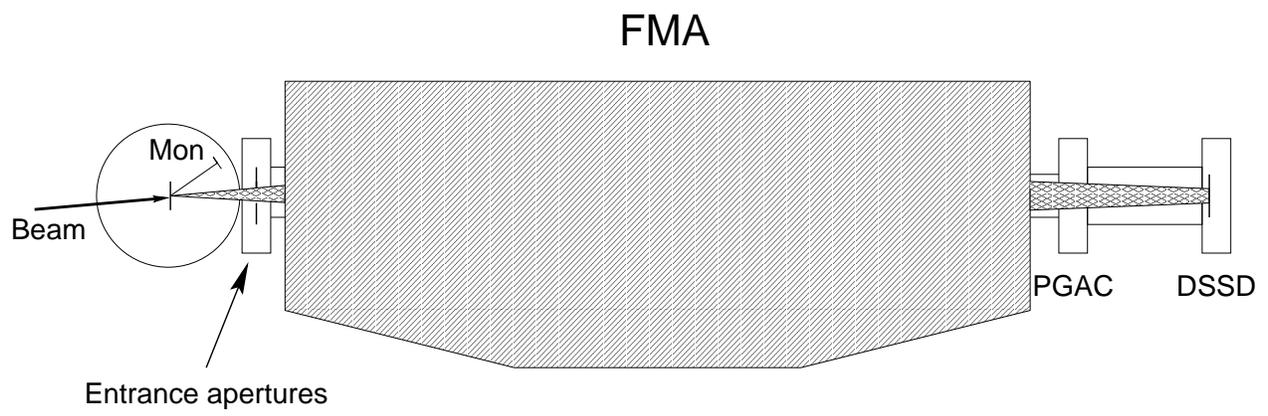
\begin{figure}[htb]
  \caption[FMA schematic]
  {Schematic illustration of the Argonne Fragment Mass
    Analyzer. The beam enters from the left.}
\label{fig1}
\end{figure}    

\begin{figure}[htb]
  \caption[FMA acceptance]
  {Separation of evaporation residues from scattered beam particles on
  the basis of time-of-flight and energy.
    }
  \label{fig2}
\end{figure}    

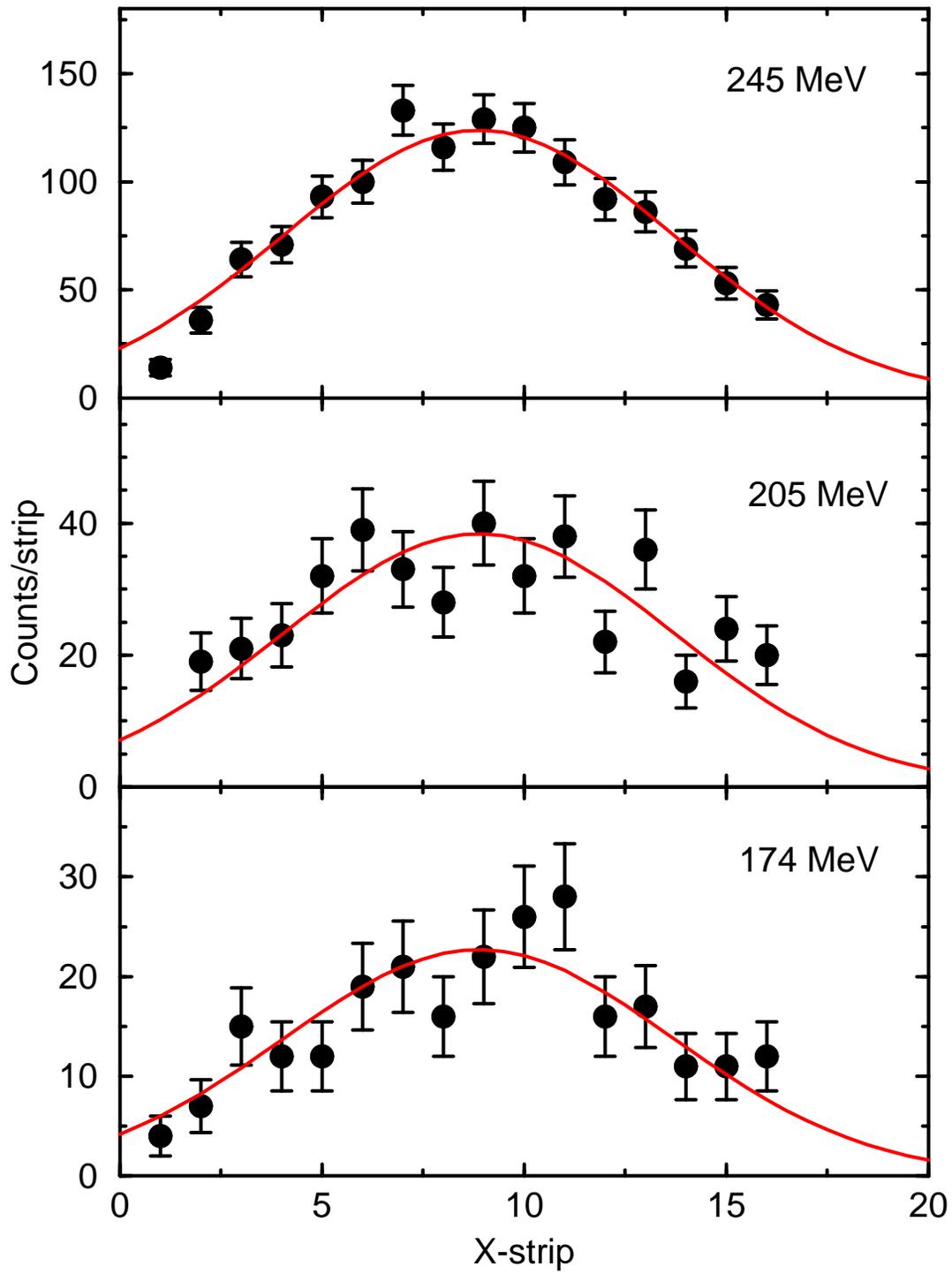
\begin{figure}[htb]
\caption[FMA efficiency and charge state]
  {The horizontal position distribution on the Si-strip detector is shown
for three beam energies. The solid curves represent Gaussian fits to the
data, from which the detection efficiency is estimated.}
\label{fig3}
\end{figure}    

\begin{figure}[htb]
\caption[S+W q-distribution]
  {Charge-state distribution, $P(q)$, of evaporation residues from the
 $^{32}$S+$^{184}$W reaction measured at three beam energies.}
\label{fig4}
\end{figure}

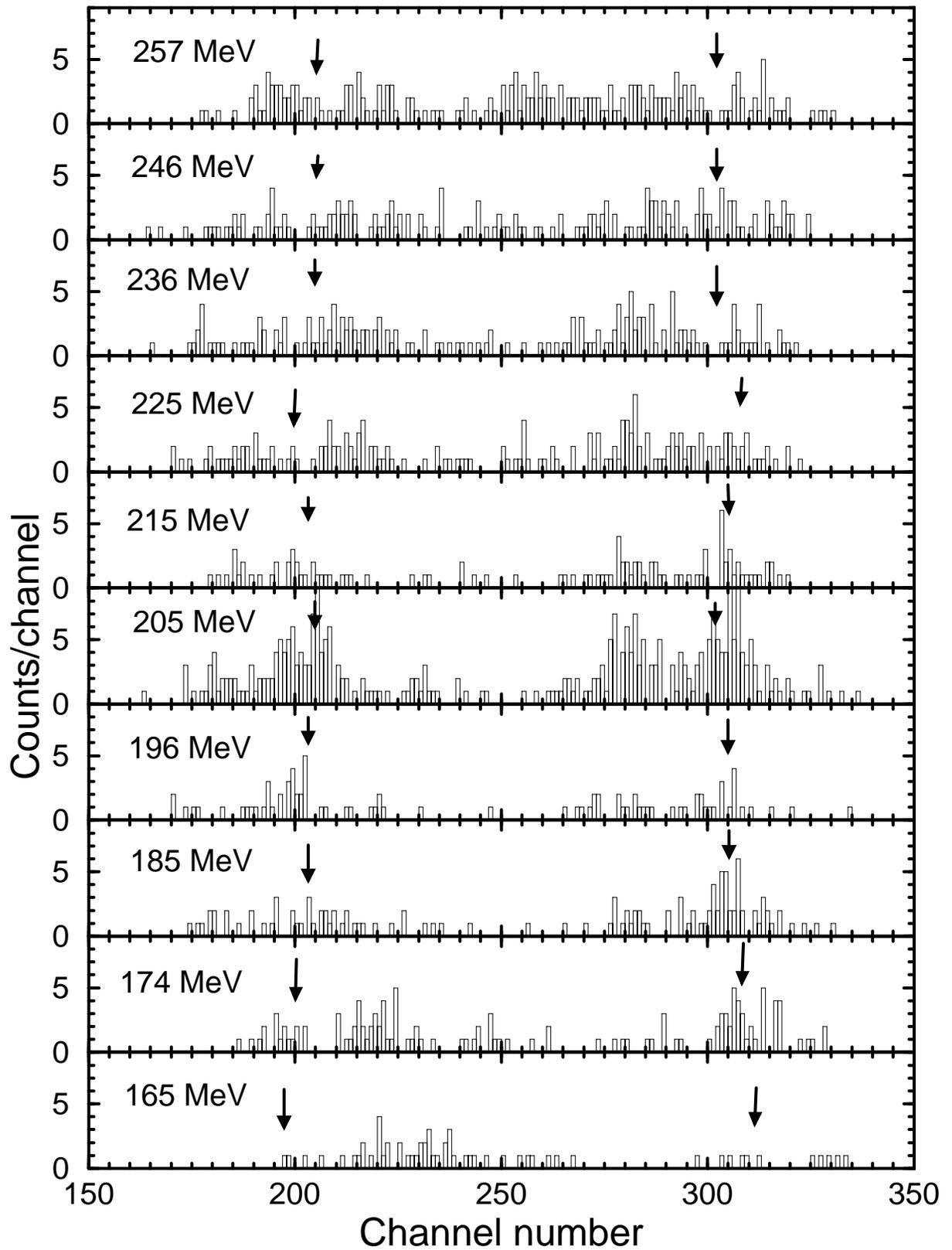
\begin{figure}[htb]
  \caption[Horizontal position]
  {Horizontal position distribution of evaporation residues along the
  focal plane, which scales with $M/q$. The arrows indicate the region of
  integration from $M/(q+\frac{1}{2})$ to $M/(q-\frac{1}{2})$.}
\label{fig5}
\end{figure}    

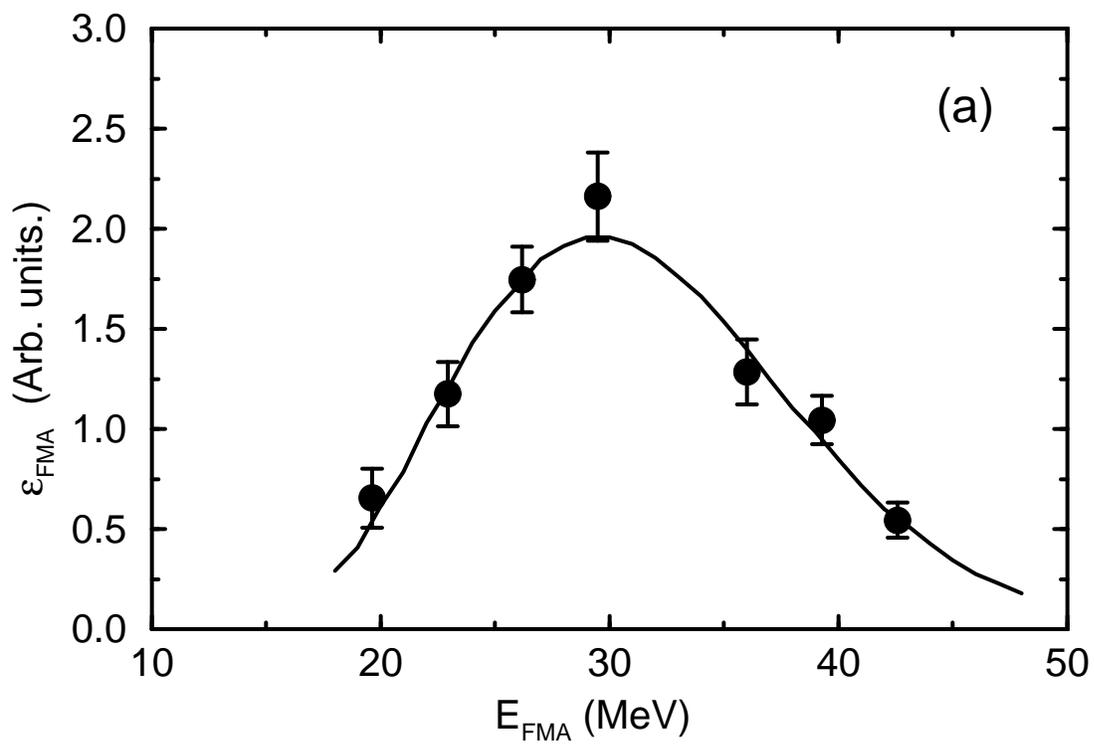
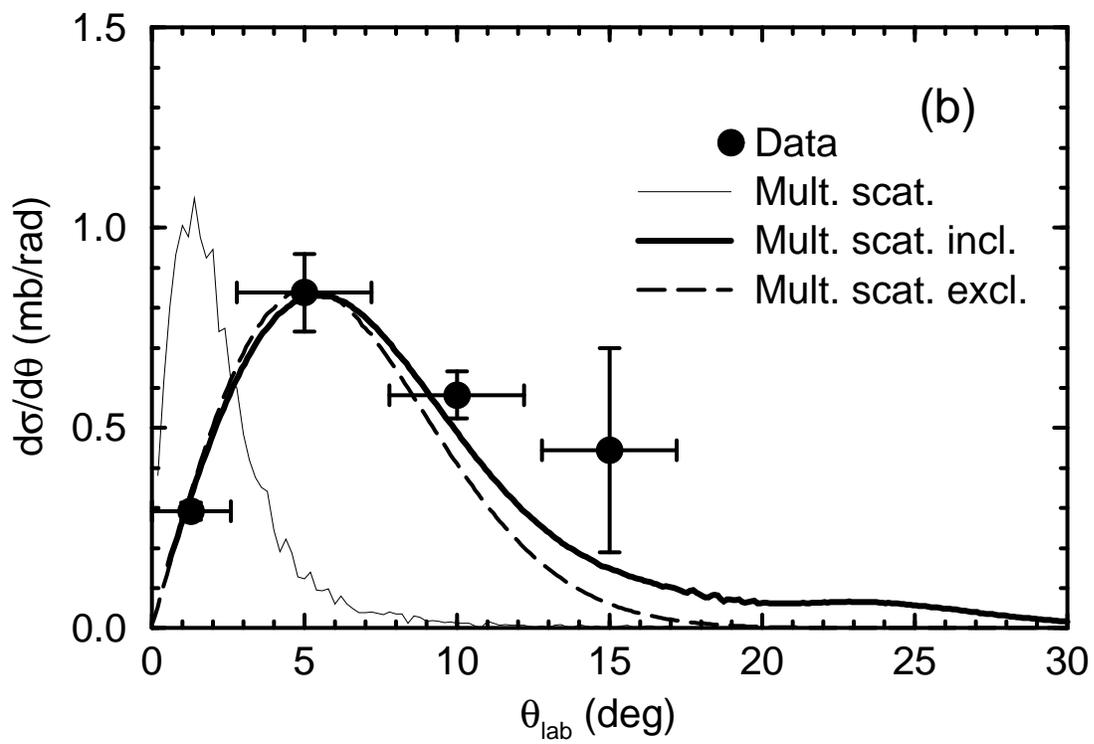
\begin{figure}[htb]
  \caption[S+W angular-distribution]
  {Top panel a: Transport efficiency $\epsilon_{FMA}$ (arbitrary units)
  for evaporation residues from 245 MeV  $^{32}$S+$^{184}$W measured with
  the FMA optimized for $M$ = 208 and $q$ = 19 and different energy $E_{FMA}$
  settings (solid points).  The solid curve results from a calculation
detailed in the text. Bottom panel b: Angular distribution
$d\sigma/d\theta_{lab}$ for the same reaction is shown as solid points
(assuming a transport efficiency of $\epsilon$ = 1.0 to the focal plane. The
thick solid and dashed curves represent calculations to fit simultaneously
the transport efficiency data (panel a) and the  angular distributions with
and without multiple scattering in target and reset foils, respectively. The
thin curve shows the calculated multiple scattering distribution.}

\label{fig6}
\end{figure}    

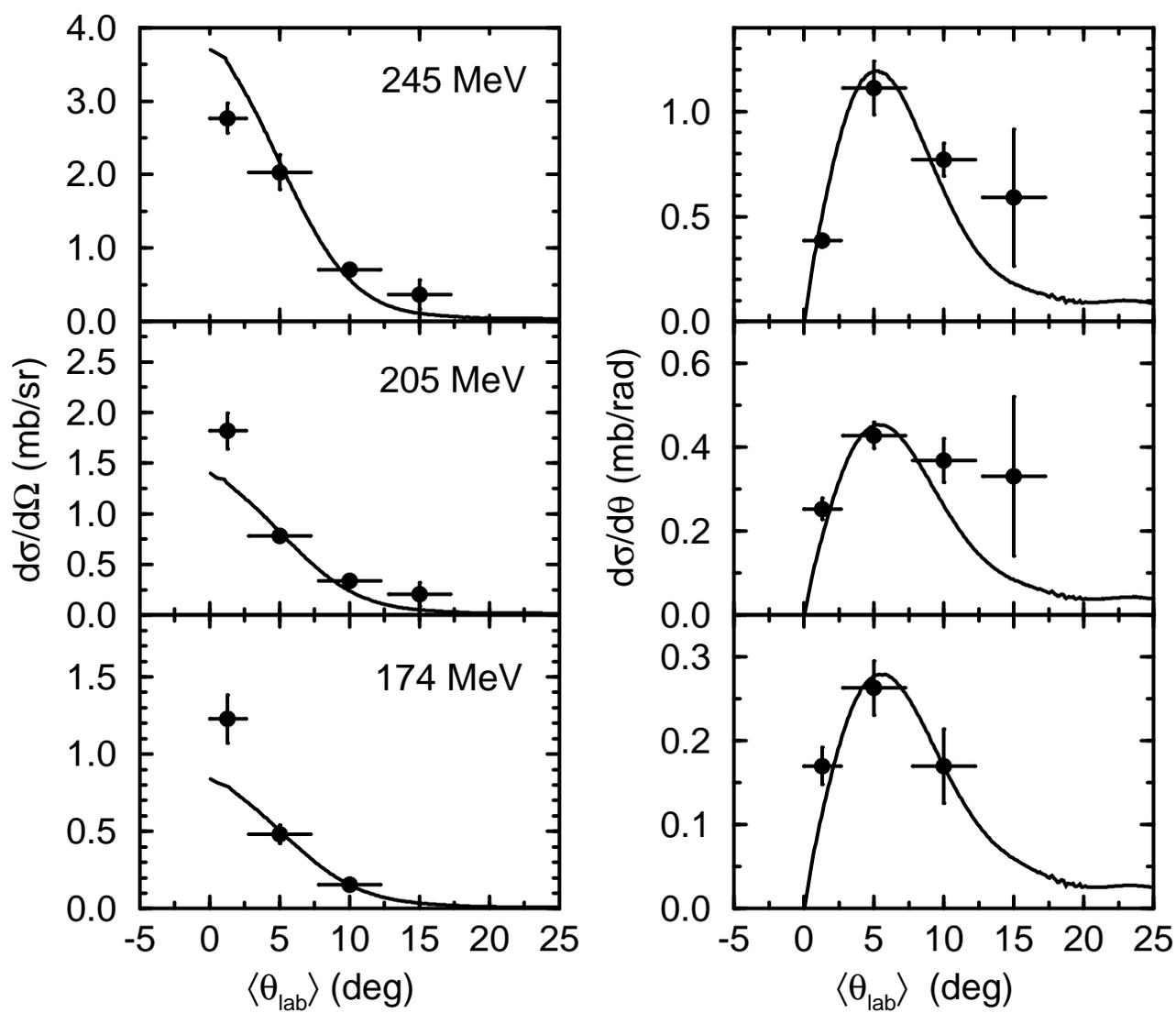
\begin{figure}[htb]
  \caption[S+W angular-distribution]
  {Right panels: Experimental angular distributions, $d\sigma/d\theta_{lab}$
  are compared with calculations (solid curves), see text, at three beam
  energies. Left panels: the same data and curves are shown in a
$d\sigma/d\Omega$ representation.}
\label{fig7}
\end{figure}

\begin{figure}[htb]
   \caption[Cross sections]
   {Evaporation residue cross sections for the reaction $^{32}$S+$^{184}$W
   (solid points) are compared with statistical model calculations with  
   and without fission hindrance (see text). The 
   capture cross section (fission + quasi-fission) 
   (open circles) and estimates of the complete fusion (open squares) cross 
   sections are shown in comparison with theoretical calculations (short 
   dashed and long-dashed curves), see text. The latter calculation were
   used to provide the initial spin distribution for the statistical model
   calculation of the evaporation residue cross section.}
\label{fig8}
\end{figure}

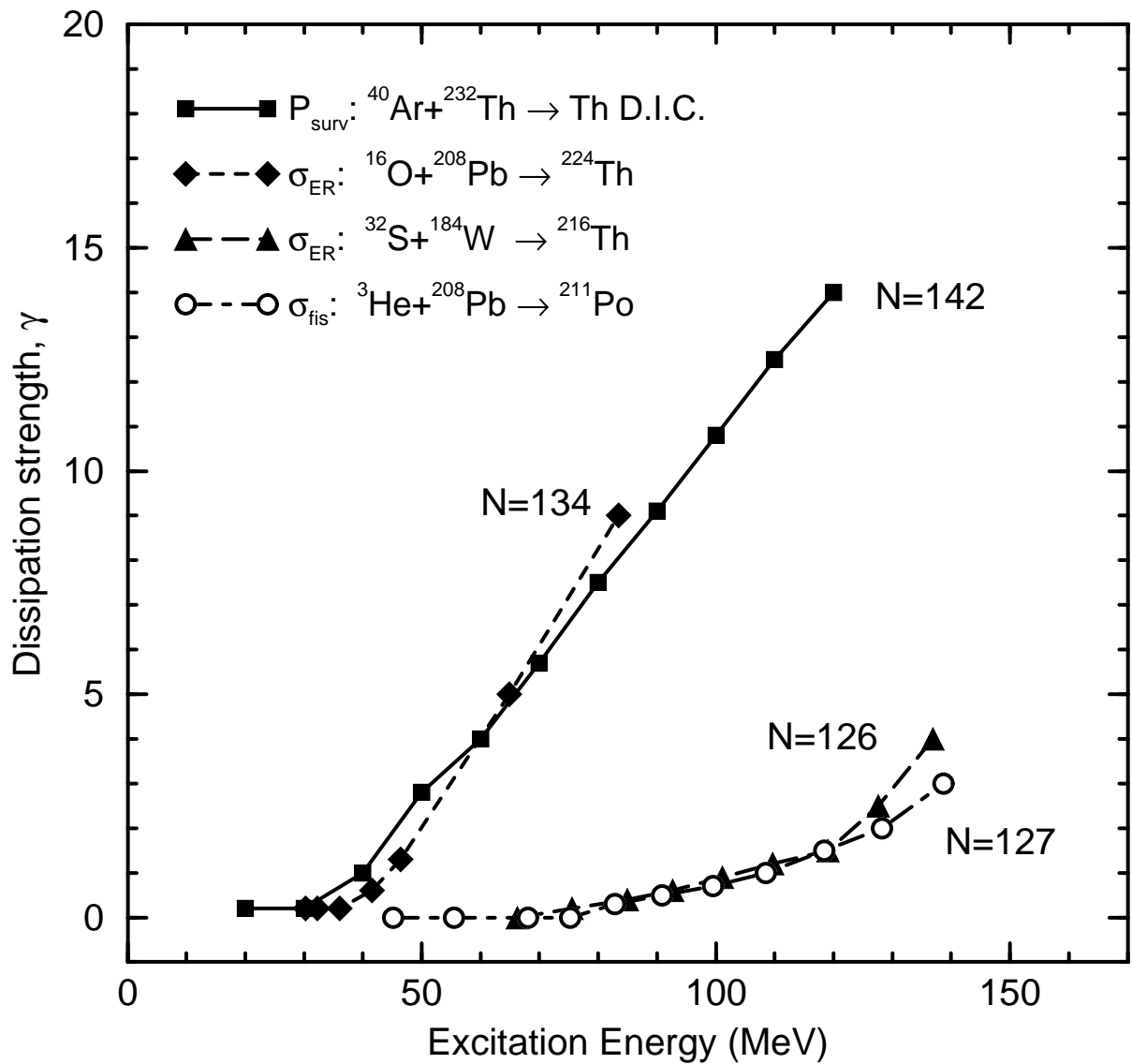
\begin{figure}[htb]
   \caption[Dissipation strength]
   { Comparison of the fission dissipation strength, $\gamma$, required to
   to reproduce different data.}
\label{fig9}
\end{figure}
}

\vspace{0.375in}

\begin{thebibliography}{99}

\bibitem[1]{Hinde}
	D. Hinde {\it et al.}, Phys. Rev. {\bf C45}, 1229 (1992)
\bibitem[2]{Hofman}
	D. Hofman {\it et al.}, Phys. Rev. Lett. {\bf 72}, 470 (1994)
\bibitem[3]{Lestone}
	J. Lestone {\it et al.}, Phys. Rev. Lett. {\bf 70}, 2245 (1993)
\bibitem[4]{Froebrich}
	P. Fr\"{o}brich {\it et al.}, Nucl. Phys. {\bf A556}, 281 (1993)
\bibitem[5]{Davids}
	C. N. Davids {\it et al.}, Nucl. Inst. Meth. {\bf B70}, 358 (1992)
\bibitem[6]{Back}
        B. B. Back {\it et al.}, Nucl. Inst. Meth. {\bf A379}, 206 (1996)
\bibitem[7]{Wollnik}
	H. Wollnik {\it et al.}, Nucl. Inst. Meth. {\bf A258}, 408 (1987)
\bibitem[8]{Shima}
	K. Shima {\it et al.}, Nucl.Inst. Meth {\bf 200}, 605 (1982)
\bibitem[9]{Glagola}
	B. G. Glagola {\it et al.}, Phys. Rev. {\bf C29}, 486 (1984)
\bibitem[10]{Keller}
	J. G. Keller {\it et al.}, Phys. Rev. {\bf C36}, 1364 (1987)
\bibitem[11]{Toke}
	J. T\={o}ke {\it et al.}, Nucl. Phys.  {\bf A440}, 327 (1985)
\bibitem[12]{Sierk}
	A. Sierk, Phys. Rev. {\bf C33}, 2039 (1986)
\bibitem[13]{Toke81}
        J. T\={o}ke and W. J. Swiatecki, Nucl. Phys. {\bf A372}, 141 (1981)
\bibitem[14]{Ignatyuk}
        A. V. Ignatyuk {\it et al}, Nucl. Phys. {\bf A593}, 519 (1995);
        Yad. Fiz. {\bf 21}, 1185 (1975), [Sov. J. Nucl. Phys. {\bf 21},
	612 (1975)]
\bibitem[15]{Hofman98}
        D. J. Hofman {\it et al},
	submitted to Phys. Rev. Lett. 
\bibitem[16]{Bohr-Wheeler}
        N. Bohr and J. A. Wheeler,
	Phys. Rev. {\bf 56}, 426 (1939)
\bibitem[17]{Kramers}
        H. A. Kramers,
	Physica {\bf 7}, 284 (1940)
\bibitem[18]{Grange}
        P. Grang\'e,Li Jun-Qing, and H. A. Weidenm\"uller,
	Phys, Rev. C{\bf 27}, 2063 (1983)
\bibitem[19]{Butsch}
        R. Butsch, D. J. Hofman, C. P. Montoya, P. Paul, and M. Thoennessen,
	Phys. Rev. C{\bf 44}, 1515 (1991)
\bibitem[20]{Back93}
        B. Back,
	in Proceedings of the International School-Seminar on Heavy Ion Physics,
	Dubna, Russia, May 10-15, 1993, ed. Y. Oganessyan, Dubna Press, 1993.
\bibitem[21]{Blocki}
        J. B\/locki, Y. Boneh, J. R. Nix, J. Randrup, M. Pobel, A. J. Sierk,
	and W. J. Swiatecki,
	Ann. Phys. (N.Y.) {\bf 113}, 330 (1978)
\bibitem[22]{Brinkmann}
        K.-T. Brinkmann {\it et al.},
	Phys. Rev. C {\bf 50}, 309 (1994)
\bibitem[23]{Hofman96}
        D. J. Hofman, B. B. Back, and P. Paul,
	Nucl. Phys. {\bf A599}, 23c (1996)
\bibitem[24]{Rubehn}
        Th. Rubehn {\it et al}, 
	Phys. Rev. C {\bf 54}, 3062 (1996)
\bibitem[25]{Back98}
        B. B. Back {\it et al},
	In proceedings of the ``International Conference on Fission
	and Properties of Neutron-rich Nuclei'', Sanibel Island, Nov 1997,
	World Scientific, 1998.
\bibitem[26]{Froebrich}
         P. Fr\"obrich and I. I. Gontchar,
	 Phys. Rep. {\bf 292}, 131 (1998)
\bibitem[27]{Dhara}
         A. K. Dhara, K. Krishan, C. Bhattacharya, and S. Bhattacharya,
	 Phys. Rev. C{\bf 57}, 2453 (1998)
\bibitem[28]{Hofmann}
         H. Hofman, F. A. Ivanyuk, and S. Yamaji,
	 Nucl. Phys. A{\bf 598}, 187 (1996)
\bibitem[29]{Kolomietz}
         V. M. Kolomietz, V. A. Plujko, and S. Shlomo,
	 Phys. Rev. C{\bf 54}, 3014 (1996)
\bibitem[30]{Magierski}
        P. Magierski, J. Skalski, and J. Blocki,
	Phys. Rev. C{\bf 56}, 1011 (1997)
\bibitem[31]{Mukhopadhyay}
        T. Mukhopadhyay and S. Pal,
	Phys. Rev. C{\bf 56}, 296 (1997)
\end{thebibliography}
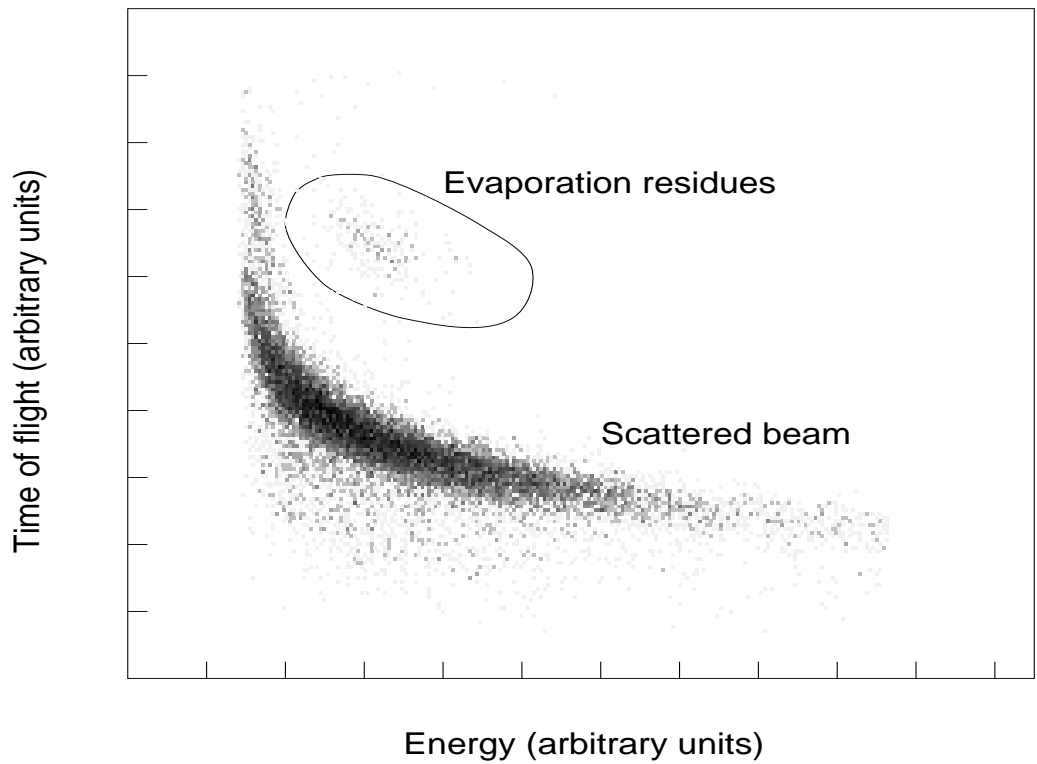
\end{document}